\documentclass[prc,twocolumn,nofootinbib,superscriptaddress,showpacs,aps]{revtex4-1}

\usepackage{amsmath}
\usepackage{amssymb}
\usepackage{graphicx}
\usepackage{dcolumn}
\usepackage{bm,braket}
\usepackage{hyperref}
\usepackage{isotope}
\usepackage{color}
\usepackage{multirow}
\usepackage[activate={true,nocompatibility},final,tracking=true,
kerning=true,spacing=true,factor=1100,stretch=10,shrink=10]{microtype}

\usepackage[all]{nowidow}
\hypersetup{
     colorlinks   = true,
     linkcolor = blue,
     citecolor = blue,
     menucolor = black,
     urlcolor = blue  
}

\begin{document}

\newcommand{\la}{\langle}
\newcommand{\ra}{\rangle}
\newcommand{\tl}{\widetilde{L}}
\newcommand{\tlp}{\widetilde{L'}}
\newcommand{\tj}{\widetilde{J}}

\renewcommand{\vec}[1]{\mathbf{#1}}
\newcommand{\clebschG}[6]{\mathcal{C}_{#1 #2 #3 #4}^{#5 #6}}  
\newcommand{\kf}{k_{\rm F}}
\newcommand{\clebschGA}[6]{
  \left(
  \begin{array}{cc|c}
  #1 & #3 & #5 \\
  #2 & #4 & #6
  \end{array}
  \right)
}
\newcommand{\sixJsymbsixA}[6]{
	\begin{Bmatrix}
  #1 & #3 & #5 \\
  #2 & #4 & #6
	\end{Bmatrix}
}  
\newcommand{\nineJsymbA}[9]{
	\begin{Bmatrix}
  #1 & #4 & #7 \\
  #2 & #5 & #8\\
  #3 & #6 & #9 
	\end{Bmatrix}
}  

\title{Asymmetric nuclear matter based on chiral two- and three-nucleon interactions}

\author{C.\ Drischler}
\email[Email:~]{christian.drischler@physik.tu-darmstadt.de}
\affiliation{Institut f\"ur Kernphysik, 
Technische Universit\"at Darmstadt, 64289 Darmstadt, Germany}
\affiliation{ExtreMe Matter Institute EMMI, 
GSI Helmholtzzentrum f\"ur Schwerionenforschung GmbH, 64291 Darmstadt, Germany}

\author{K.\ Hebeler}
\email[Email:~]{kai.hebeler@physik.tu-darmstadt.de}
\affiliation{Institut f\"ur Kernphysik, 
Technische Universit\"at Darmstadt, 64289 Darmstadt, Germany}
\affiliation{ExtreMe Matter Institute EMMI, 
GSI Helmholtzzentrum f\"ur Schwerionenforschung GmbH, 64291 Darmstadt, Germany}

\author{A.\ Schwenk}
\email[Email:~]{schwenk@physik.tu-darmstadt.de}
\affiliation{Institut f\"ur Kernphysik, 
Technische Universit\"at Darmstadt, 64289 Darmstadt, Germany}
\affiliation{ExtreMe Matter Institute EMMI, 
GSI Helmholtzzentrum f\"ur Schwerionenforschung GmbH, 64291 Darmstadt, Germany}

\begin{abstract}
We calculate the properties of isospin-asymmetric nuclear matter based
on chiral nucleon-nucleon (NN) and three-nucleon (3N) interactions. To
this end, we develop an improved normal-ordering framework that allows
to include general 3N interactions starting from a plane-wave
partial-wave-decomposed form. We present results for the energy per
particle for general isospin asymmetries based on a set of different
Hamiltonians, study their saturation properties, the
incompressibility, symmetry energy, and also provide an analytic
parametrization for the energy per particle as a function of density
and isospin asymmetry.
\end{abstract}


\maketitle

\section{\label{sec:intro}Introduction}

Microscopic calculations of isospin-asymmetric nuclear matter are
important for nuclear physics and astrophysical applications. They
allow to give ab initio constraints for key quantities for our
understanding of core-collapse supernovae and neutron stars. In
addition, they can guide energy-density functionals for global
predictions of nuclear properties.

Advances in chiral effective field theory (EFT)~\cite{Epel09RMP,mach11pr} and
renormalization group methods~\cite{Bogn09PPNP,Furn13RPP} have opened
the way to improved and systematic studies of nuclear matter as well
as finite nuclei~\cite{hamm12rmp,Hebe15ARNPS}. For symmetric matter it
was found that low-momentum NN plus 3N interactions are capable to
predict realistic saturation properties, with 3N forces fit only to
few-body data~\cite{Hebe11fits}, whereas neutron matter was found to
be perturbative~\cite{Hebe10nmatt}. In subsequent studies, symmetric
matter and neutron matter were also investigated based on chiral EFT
interactions within the self-consistent Green's function
framework~\cite{Carb13SymE,Carb14SCGF}, using coupled-cluster
theory~\cite{hage13ccnm,ekst15AccNuc}, with in-medium chiral perturbation
theory~\cite{Holt13PPNP} and in many-body perturbation
theory~\cite{Cora14MBPT}. Furthermore, the development of novel local
chiral NN forces opened the way to first Quantum Monte Carlo
studies of neutron matter based on chiral EFT
interactions~\cite{Geze13QMCchi,Geze14long,tews15QMC3N,lynn15liNuc3N}. The results of these
studies also represent first nonperturbative validation of
many-body perturbation theory for neutron matter.

Asymmetric nuclear matter has been studied within various many-body
approaches during the last decades based on phenomenological NN
potentials~\cite{Brue68Sym,Laga81Varia,Bomb91ANM,Zuo99ANM,
Zuo02ANM,Vida09ESym,Fric05ANM}. Chiral EFT interactions have been
applied to asymmetric matter only
recently~\cite{Dris14asymmat,Well15ThermANM,Kaise15S4}.  Explicit
calculations at general proton fractions allow to extract key
quantities like the nuclear symmetry energy more microscopically,
because no empirical parametrizations for the energy as a function of
the isospin asymmetry are needed. Commonly, such parametrizations were
either based on the standard quadratic expansion (see, e.g.,
Ref.~\cite{Well15ThermANM} for a recent work) or inspired by the form
of energy-density functionals~(see, e.g.,~Ref.~\cite{Hebe13ApJ}).

A major challenge for performing such many-body calculations is the
treatment of 3N forces and the quantification of theoretical
uncertainties. In contrast to many-body uncertainties, which can be
investigated by benchmarking, the quantification of uncertainties in
the nuclear Hamiltonian is a more challenging task (see e.g.,
Ref.~\cite{furn15truncErr}). There are currently ongoing efforts to
develop novel chiral EFT interactions (see, e.g.,
Ref.~\cite{Epel14improved,Epel14NNn4lo,carl15UncAna}) that enable order-by-order
studies of matter and nuclei in the chiral expansion and allow to test
the validity of the chiral power counting at nuclear densities in a
systematic way.

For these investigations 3N forces play a central role. In Weinberg
power counting the leading 3N forces at N$^2$LO contain two unknown
low-energy couplings $c_D$ and $c_E$, whereas the subleading 3N forces
at N$^3$LO do not contain any new low-energy
couplings~\cite{Bern083Nlong,Bern113Nshort}.  First full N$^3$LO
calculations of neutron matter showed that subleading 3N forces at
N$^3$LO provide significant contributions to the energy per
particle~\cite{Tews13N3LO,krue13n3lolong}. This could be an indication
for a slow convergence of the chiral expansion for 3N forces. These
findings were confirmed by first explorative calculations of symmetric
matter up to N$^3$LO~\cite{krue13n3lolong}. Due to the complexity and
rich analytical structure of 3N forces at
N$^3$LO~\cite{Bern083Nlong,Bern113Nshort,Hebe12msSRG} the 3N
contributions at this order could only be included in the Hartree-Fock
approximation in these studies. While this approximation is expected
to be reasonable for neutron matter, such a treatment is certainly not
reliable for sufficiently large proton fractions and consequently
higher-order many-body contributions need to be included.  For the
same reason the calculations of asymmetric nuclear matter of
Ref.~\cite{Dris14asymmat} were limited to small proton fractions.

In this paper, we present a framework that allows to include general
3N forces in calculations of asymmetric nuclear matter systematically
and hence allows to extend the studies of Ref.~\cite{Dris14asymmat} to
arbitrary proton fractions. Our calculations are based on a set of
seven Hamiltonians with NN interactions at N$^3$LO evolved with the
similarity renormalization group (SRG) to different resolution scales
$\lambda$ plus 3N interactions at N$^2$LO with 3N cutoff
$\Lambda_{\text{3N}}$:
\begin{equation}
H(\lambda,\Lambda_{\text{3N}}) = 
T + V_\text{NN}(\lambda) + V_\text{3N}(\Lambda_{\text{3N}}) \, .
\end{equation}
By using 3N forces at N$^2$LO as a truncated basis and assuming the
long-range couplings $c_i$ to be invariant under the SRG
transformation, the 3N short-range couplings $c_D$, $c_E$ were fit in
Ref.~\cite{Hebe11fits} for seven combinations of $\lambda /
\Lambda_{\text{3N}}$ to the experimental binding energy of $^3$H and
the charge radius of $^4$He. The resulting values of the low-energy
couplings are listed in Table~\ref{tab:couplings}. This set of
Hamiltonians serves as an estimate for the theoretical uncertainties
due to nuclear forces in our many-body calculations.

\begin{table*}[t]
\caption{\label{tab:couplings}
The set of seven Hamiltonians used for the many-body calculations of this
study. The low-energy couplings $c_D$, $c_E$ were fit in Ref.~\cite{Hebe11fits}
to the binding energy of $^3$H and the charge radius of $^4$He for
given SRG resolution scale $\lambda$, 3N cutoff $\Lambda_\text{3N}$
and the long-range couplings $c_i$. The Hamiltonians are based on the
N$^3$LO NN potential EM~500~MeV~\cite{Ente03EMN3LO}, except for
Hamiltonian~6*, which is based on the N$^3$LO NN potential
EGM~550/600~MeV~\cite{Epel05EGMN3LO}. Moreover, the Hamiltonians use
consistent $c_i$ values in NN and 3N interactions, except for
Hamiltonian~7, which uses the $c_i$ values from the NN partial-wave
analysis of Ref.~\cite{Rent03ciPWA} in 3N interactions. We refer to
Sec.~\ref{sec:Disc_EOS} for a discussion of the special treatment of
Hamiltonian~6*.}
\begin{ruledtabular}
\begin{tabular}{llcccccc} 
& NN Potential & $\lambda/\Lambda_{\text{3N}}$ [fm$^{-1}$] & $c_1$~[GeV$^{-1}$] & $c_3$~[GeV$^{-1}$] &$c_4$~[GeV$^{-1}$] & $c_D$ & $c_E$ \\
\hline
\#1 &EM~500~MeV& $1.8/2.0$ & $-0.81$ & $-3.2$ & $5.4$ & $1.264$ & $-0.120$ \\
\#2 &EM~500~MeV& $2.0/2.0$ & $-0.81$ & $-3.2$ & $5.4$ & $1.271$ & $-0.131$ \\
\#3 &EM~500~MeV& $2.0/2.5$ & $-0.81$ & $-3.2$ & $5.4$ & $-0.292$ & $-0.592$ \\
\#4 &EM~500~MeV& $2.2/2.0$ & $-0.81$ & $-3.2$ & $5.4$ & $1.214$ & $-0.137$ \\
\#5 &EM~500~MeV& $2.8/2.0$ & $-0.81$ & $-3.2$ & $5.4$ & $1.278$ & $-0.078$ \\
\#6* &EGM~550/600~MeV& $2.0/2.0$ & $-0.81$ & $-3.4$ & $3.4$ & $-4.828$ & $-1.152$ \\
\#7 &EM~500~MeV& $2.0/2.0$ & $-0.76$ & $-4.78$ & $3.96$ & $-3.007$ & $-0.686$ 
\end{tabular}
\end{ruledtabular}
\end{table*}

The paper is organized as follows. In Sec.~\ref{sec:normal-ordering},
we introduce an improved density-dependent NN interaction to include
3N-force contributions in our calculations. In Sec.~\ref{sec:MB-calc},
we discuss the expressions for the energy per particle to first and
second order in many-body perturbation theory for general isospin
asymmetries. In Sec.~\ref{sec:results}, we present our microscopic
results for the energy per particle for eleven proton fractions based
on a set of different Hamiltonians. We study their saturation
properties, the incompressibility, symmetry energy, and also provide
an analytic global fit of our results. Finally, we conclude and
summarize in Sec.~\ref{sec:summary}.

\section{\label{sec:normal-ordering}Improved normal ordering}

Normal ordering is a key step for the practical treatment of 3N forces
as effective two-body interactions in many-body calculations of matter
and nuclei. It allows to rewrite the 3N-force part of the Hamiltonian
exactly in terms of normal-ordered zero-, one- and two-body
contributions plus a residual three-body term (for details see
Ref.~\cite{Bogn09PPNP}). In infinite matter normal ordering involves
a summation of one particle over occupied states in the Fermi sphere
(see also Refs.~\cite{Holt10ddNN,Hebe10nmatt}). For 3N forces this
summation can be expressed formally in the form:
\begin{equation}\label{eq:normord_singpart}
\overline{V}_{\rm{3N}} =
\text{Tr}_{\sigma_3} \text{Tr}_{\tau_3} \int \frac{d \mathbf{k}_3}{(2 \pi)^3}
\, n_{\mathbf{k}_3}^{\tau_3} \, \mathcal{A}_{123} V_{\rm{3N}} \, ,
\end{equation}
which involves sums over spin and isospin projection quantum numbers
$\sigma_3$ and $\tau_3$ as well as an integration over all momentum
states, weighted by the momentum distribution functions
$n^{\tau_3}_{\bf k}$ for a given neutron and proton density. In the
following, we choose the Fermi-Dirac distribution function at zero
temperature, $n^{\tau_3}_{\bf k} = \theta(\kf^{\tau_3} - |{\bf k}|)$,
and we assume spin-unpolarized and homogeneous matter. We can apply
the present framework also to general correlated distributions
functions. However, it was shown in infinite matter~\cite{Carb14SCGF}
that the energy is not very sensitive to the particular choice of the
reference state for the chiral EFT interactions used in this
work. This indicates that the residual 3N contributions are very small
such that they can be neglected. $V_{\rm{3N}}$ represents the 3N
interaction, whereas $\mathcal{A}_{123}$ is the three-body
antisymmetrizer. The effective interaction $\overline{V}_{\rm{3N}}$ in
Eq.~\eqref{eq:normord_singpart} represents a density-dependent NN
interaction that can be combined with contributions from free-space NN
interactions.

The 3N interaction $V_{\rm{3N}}$ is the fundamental microscopic input
to Eq.~\eqref{eq:normord_singpart}. The momentum dependence of a
general translationally invariant 3N interaction can be most
efficiently expressed as a function of the Jacobi momenta
\begin{equation} 
\mathbf{p} = \frac{\mathbf{k}_1 - \mathbf{k}_2}{2}\, , \quad \mathbf{q} 
= \frac{2}{3} \left[ \mathbf{k}_3 - \frac{1}{2} (\mathbf{k}_1 
+ \mathbf{k}_2) \right],
\end{equation} 
where $\mathbf{k}_i$ denote the single-nucleon momenta. In the
following $\mathbf{p}$ and $\mathbf{q}$ ($\mathbf{p}'$ and $\mathbf{q}'$)
denote the Jacobi momenta of the initial (final) state: 
\begin{equation}
V_{\rm{3N}} = V_{\rm{3N}} (\mathbf{p},\mathbf{q},\mathbf{p}',\mathbf{q}') \, .
\end{equation}
Hence, it is natural to perform the normal ordering
Eq.~\eqref{eq:normord_singpart} in this Jacobi basis. By expressing
all single-particle momenta in terms of the Jacobi momenta and the
two-body center-of-mass momentum $\mathbf{P} = \mathbf{k}_1 +
\mathbf{k}_2 = \mathbf{k}'_1 + \mathbf{k}'_2$ we obtain:
\begin{equation}
\overline{V}_{\rm{3N}} =  \left( \frac{3}{2}\right)^3 \text{Tr}_{\sigma_3}
\text{Tr}_{\tau_3} \int \frac{d \mathbf{q}}{(2
\pi)^3} \, n_{(3 \vec{q}+\vec{P})/2}^{\tau_3} \, \mathcal{A}_{123} V_{\rm{3N}}\, .
\label{eq:normord_jacobi}
\end{equation}

The calculation of the effective interaction $\overline{V}_{\rm{3N}}$
is challenging due to the complex structure of general 3N
interactions.  For the practical treatment it is common to decompose
3N interactions in a $Jj$-coupled 3N partial-wave momentum basis of
the form~\cite{Gloe83QMFewBod,Skib11Tucson}:
\begin{equation}
\left| p q \alpha \right> \hspace{-1.6mm} \phantom \rangle \equiv \left| p q; \left[ (L S) J (l s) j \right] \mathcal{J} (T t) 
\mathcal{T} \right> \, .  \label{eq:Jj_bas}
\end{equation}
Here, $L$, $S$, $J$, and $T$ denote the relative orbital angular
momentum, spin, total angular momentum, and isospin of particles 1 and
2 with relative momentum $p$. The quantum numbers $l$, $s=1/2$, $j$
and $t=1/2$ label the orbital angular momentum, spin, total angular
momentum and isospin of particle $3$ relative to the center-of-mass
motion of particle 1 and 2. The 3N quantum numbers $\mathcal{J}$ and
$\mathcal{T}$ define the total 3N angular momentum and isospin (for
details see Ref.~\cite{Gloe83QMFewBod}). In particular, 3N
interactions do not depend on the projection quantum numbers
$m_{\mathcal{J}}$ and for isospin-symmetric interactions also not on
$m_{\mathcal{T}}$, hence we omit these labels in the basis states.

We evaluate Eq.~\eqref{eq:normord_jacobi} in this partial-wave basis. 
The basic ingredient of our normal-ordering framework are antisymmetrized
3N matrix elements of the form 
\begin{multline} \label{eq:TBF_ME}
\left< p q \alpha| V^{\text{as}}_{\text{3N}} | p' q' \alpha' \right> \\
= \left< p q \alpha | (1 + P_{123} + P_{132}) V_{\text{3N}}^{(i)} (1 + P_{123} + P_{132}) | p' q' \alpha' \right> \,, \\
\end{multline}
where $P_{123}$, $P_{132}$ are the cyclic permutation operators of
three particles and $V_{\text{3N}}^{(i)}$ represents one Faddeev component of
the 3N interaction (see Refs.~\cite{Gloe83QMFewBod, Hebe15N3LOpw} for
details).

Previous normal-ordering frameworks for infinite matter have been
developed for a specific 3N interaction, e.g., the leading chiral 3N
interactions at N$^2$LO~\cite{Holt10ddNN,Hebe10nmatt}. This makes it
necessary to re-develop expressions for the effective interaction
$\overline{V}_{\rm{3N}}$ for each new contribution and for each
isospin asymmetry. Moreover, the treatment of more complicated 3N
interactions, e.g., the subleading chiral 3N interactions at
N$^3$LO~\cite{Bern083Nlong,Bern113Nshort} becomes very tedious. In
contrast, because the partial-wave decomposition of these 3N
interactions has been completed very recently~\cite{Hebe15N3LOpw},
these contributions can be included in the present framework without
additional efforts.

Although the effective interaction $\overline{V}_{\rm{3N}}$ is an
effective NN interaction, there are important differences to
free-space interactions: due to Galilean invariance, free-space NN
interactions can only depend on the initial and final relative momenta
$\mathbf{p}$ and $\mathbf{p}'$. Since the many-body rest frame defines
a preferred frame the effective NN interaction
$\overline{V}_{\rm{3N}}$ generally also depends on the center-of-mass
momentum $\mathbf{P}$. In particular, the interaction also depends on
the angle between the momenta $\mathbf{p}, \mathbf{p}'$ and
$\mathbf{P}$, which leads to a much more complicated partial-wave
structure than for free-space NN interactions. In order to avoid these
complications, the approximation $\mathbf{P}=0$ has been imposed for
the effective NN interaction in previous
works~\cite{Holt10ddNN,Hebe10nmatt,Carb14SCGF}.

The flexibility of the present framework allows to extend the
calculation of $\overline{V}_{\rm{3N}}$ to finite momenta
$\mathbf{P}$. In order to reduce the complexity of the effective
interaction and to simplify its application in many-body calculations
we average the direction of $\mathbf{P}$ over all angles:
\begin{equation}
n_{(3 \vec{q}+\vec{P})/2}^\tau \longrightarrow \Gamma^\tau(q,P) = \frac{1}{4 \pi}\int d\Omega_{\vec{P}} \, n_{(3\vec{q}+\vec{P})/2}^\tau \, ,
\end{equation}
with
\begin{equation}
\Gamma^\tau(q,P)
=\begin{cases} 
1 & (3q+P) \leqslant 2 k_{F,\tau} \; , \\
0 & |3q-P| \geqslant 2 k_{F,\tau} \; ,  \\
\frac{1}{2}\int_{-1}^{\gamma} d\cos  \theta \,n_{(3\vec{q}+\vec{P})/2}^\tau  & \text{else\,,}
\end{cases}
\end{equation}
and $\gamma = (4 k_{F,\tau}^2-9q^2-P^2)/(6Pq)$. Within this
approximation the effective interaction $\overline{V}_{\rm{3N}}$
acquires an additional dependence on the absolute value of
$\mathbf{P}$, whereas its partial-wave structure is still sufficiently
simple so that it can be combined with free-space NN interactions in
many-body calculations in a straightforward way.  Explicitly, we
obtain for the partial-wave matrix elements normalized to the direct
term:
\begin{align} \label{eq:V_eff_pw}
& \left< p (L S) J T m_T|\overline{V}_{\rm{3N}}^{\rm{as}} (P) | p' (L' S') J T' m_T \right> \nonumber \\
& = \frac{(-i)^{L'-L}}{(4\pi)^2} \, \left(\frac{3}{4 \pi} \right)^3 3 \int dq \, q^2\; f_\text{R}(p,q) f_\text{R}(p',q) \nonumber  \\
& \quad \times \sum \limits_{\tau}  \clebschG{T}{m_{T}}{1/2}{\tau}{\mathcal{T}}{m_{T}+\tau} \clebschG{T'}{m_{T}}{1/2}{\tau}{\mathcal{T}}{m_{T}+\tau}  \; \Gamma^\tau (q,P) \nonumber  \\
& \quad \times \sum \limits_{\substack{l,j\\\mathcal{J},\mathcal{T}}} \frac{2\mathcal{J}+1}{2 J+1} \delta_{ll'} \delta_{jj'} \delta_{JJ'} \braket{pq \alpha| V^{\text{as}}_{\text{3N}}|p'q \alpha'} \, ,
\end{align}
where $f_\text{R}(p,q)$ denotes the non-local 3N regulator
function. We will use the form $f_\text{R}(p,q)=\exp [-((p^2+ 3
q^2/4)/ \Lambda_{\text{3N}}^2)^{4}]$ following
Ref.~\cite{Hebe11fits}.  Because of the definition of the 3N matrix
elements in Eq.~\eqref{eq:TBF_ME}, our effective NN potential,
$\overline{V}_{\rm{3N}}^{\rm{as}} =
\mathcal{A}_{123}\overline{V}_{\rm{3N}}\mathcal{A}_{123}$, involves
two antisymmetrizers in contrast to the formal definition in
Eq.~\eqref{eq:normord_singpart}.

\begin{figure}[t]
\includegraphics[scale=1.0,clip=]{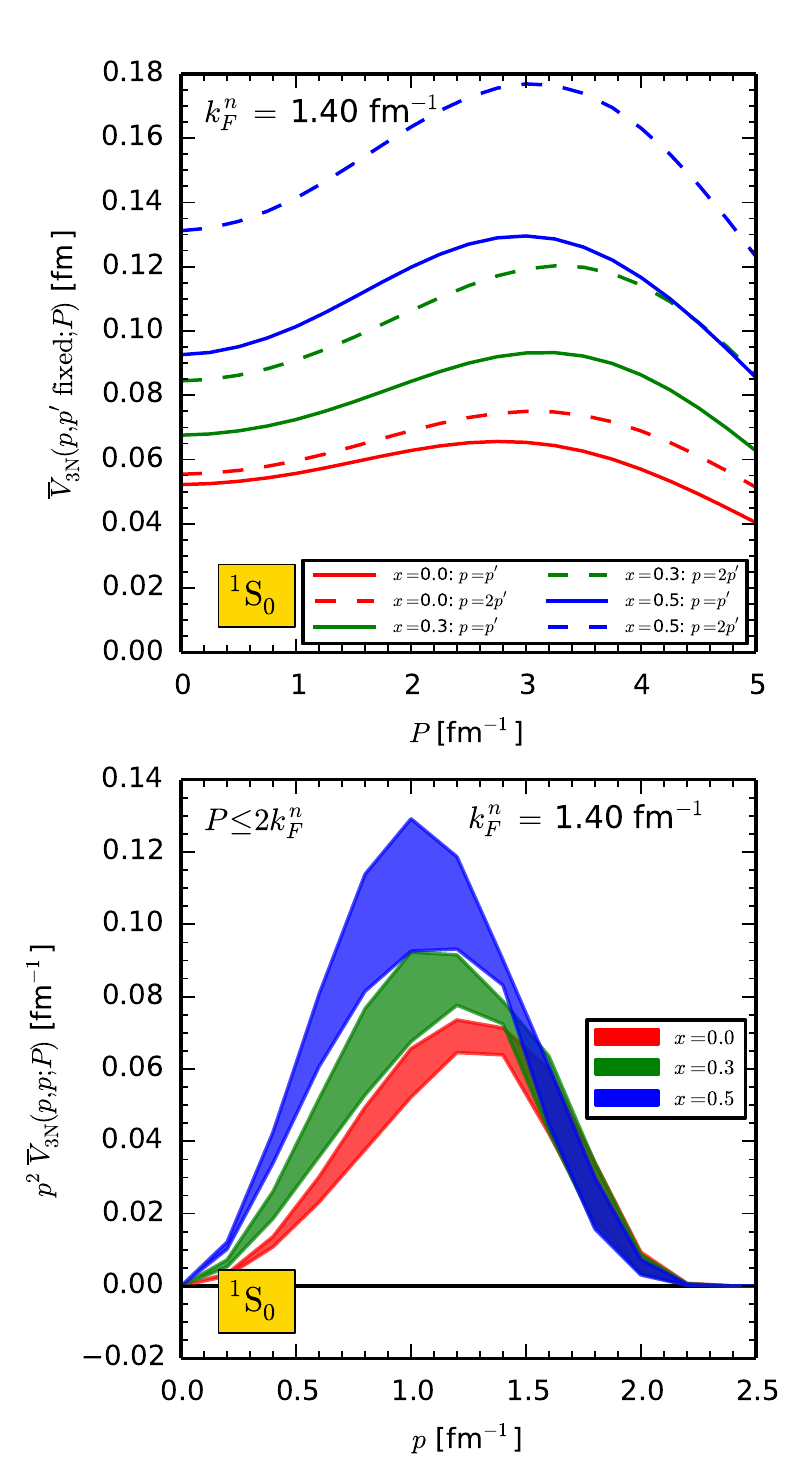}
\caption{\label{fig:Veff_me}(Color online)
The upper panel shows the matrix elements of the effective
interaction $\overline{V}_{\text{3N}} = \overline{V}_{\rm{3N}}/9$ in
the ${}^1$S$_0$ channel with $m_T=-1$ as a function of the
center-of-mass momentum $P$ for fixed relative momenta,
$p=p'=1$~fm$^{-1}$~(solid) and $p=2p'=1$~fm$^{-1}$~(dashed line), and
proton fractions $x$ at a neutron Fermi momentum $\kf^n =
1.4$~fm$^{-1}$. For the color code, see the legend in the lower
panel. The lower panel shows the diagonal matrix elements times $p^2$
as a function of the relative momentum $p$. In the first- and
second-order many-body contributions, the value of $P$ is
kinematically limited to $P \leqslant k_F^{\tau_1} + k_F^{\tau_2}$, so
for $m_T=-1$ to $P \leqslant 2k_F^n$.}
\end{figure}

\begin{figure*}[t]
\includegraphics[scale=1.0,clip=]{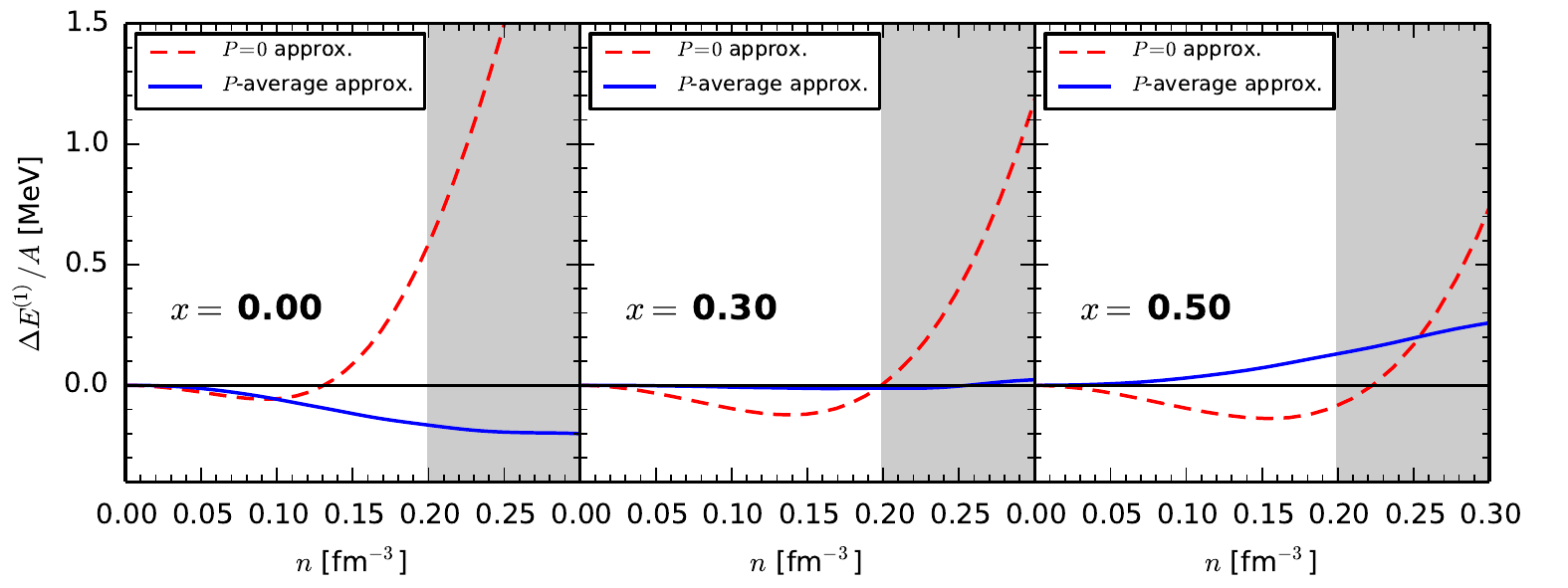}
\caption{\label{fig:comp_3N_HF}(Color online) 
Comparison of 3N Hartree-Fock energies based on Hamiltonian~2 in
Table~\ref{tab:couplings} for the $P = 0$~(red dashed) and $P$-average
approximation~(blue solid line) for the effective interaction
$\overline{V}_{\rm{3N}}^{\rm{as}}$.  Results are shown as difference
to the exact Hartree-Fock energy for three proton fractions, $x = 0$~(left), $x =
0.3$~(center), and $x = 0.5$~(right panel). The $P = 0$ values give
larger deviations above saturation density, whereas the $P$-average
approximation behaves more systematic over the entire density range.}
\end{figure*}

Note that, except for neutron and symmetry matter, off-diagonal matrix
elements in spin and isospin quantum numbers $S$ and $T$ contribute to
the effective potential. It also depends on the isospin projection
$m_T$, a direct consequence of the isospin dependence of the
occupation function $n_{k}^{\tau}$. Only in the case of neutron and
symmetric nuclear matter, the interaction is diagonal in $S$, $T$, and
also independent of the allowed $m_T$, because of isospin symmetry of
chiral 3N forces up to N$^3$LO.

In Fig.~\ref{fig:Veff_me} we present the results of some
representative matrix elements of $\overline{V}_{\text{3N}} =
\overline{V}^{\text{as}}_{\text{3N}}/9$ in the $^1$S$_0$ channel with
$m_T=-1$ ($nn$) for different proton fractions $x$ and a neutron Fermi
momentum $k_F^n=1.4$~fm$^{-1}$. The normalization of the matrix
elements is chosen such that they can be directly combined with those
of the free-space NN interaction for calculations in the Hartree-Fock
approximation. The top panel shows the matrix elements at fixed
relative momenta, $p=p'=1$~fm$^{-1}$ (solid) and $p=2p'=1$~fm$^{-1}$
(dashed line), respectively, as a function of $P$.  Due to momentum
conservation, the value of $P$ is kinematically limited to $P
\leqslant k_F^{\tau_1}+k_F^{\tau_2}$ for the first- and second-order
contributions, depending on $m_T=\tau_1+\tau_2$. The lower panel shows
the diagonal matrix elements with the measure $p^2$ as a function of
the relative momentum for this range of center-of-mass momenta. The
$P=0$ results are in excellent agreement with
Ref.~\cite{Hebe10nmatt,Hebe11fits}. For $x=0$, the matrix elements
have a rather weak dependence on $P$. This suggests that neutron
matter results can be approximated reasonably well by the $P=0$
approximation, as checked at the Hartree-Fock level in
Ref.~\cite{Hebe10nmatt}, while for increasing proton fractions the $P$
dependence of the matrix elements becomes more pronounced.

In Fig.~\ref{fig:comp_3N_HF} we compare results for the 3N
Hartree-Fock energies based on the different approximations for the
effective NN interaction. The three panels show the energy difference
to the exact Hartree-Fock result for proton fraction $x = 0$~(left),
$x = 0.3$~(center), and $x = 0.5$~(right). The effective NN
interaction based on the $P=0$ approximation reproduces the exact
results well up to $n \simeq (0.13-0.23)$~fm$^{-3}$, depending on the
proton fraction. For higher densities the deviation systematically
increases, indicating a breakdown of the $P=0$ approximation. In
contrast, the results based on the $P$-average approximation agree
well with the exact results over the entire density range.

\section{\label{sec:MB-calc}Many-body calculations}

For our many-body calculations we follow the calculational strategy of
Ref.~\cite{Dris14asymmat}. We parametrize the total density in terms
of the neutron Fermi momentum $k_\text{F}^n$ and in terms of the
proton fraction $x=n_p/n$ or, equivalently, the isospin asymmetry
$\beta=(n_n- n_p)/n=1-2x$. The neutron, proton, and total density are
labeled as $n_n$, $n_p$, and $n=n_n+n_p$, respectively. We probe the
sensitivity of our results to uncertainties of the Hamiltonian by
performing calculations for all interactions listed in
Table~\ref{tab:coeff}. These Hamiltonians start from two different NN
potentials, have different values of the SRG resolution scale
$\lambda$, different 3N cutoffs $\Lambda_{\rm{3N}}$, as well as
different values of the long-range couplings $c_i$. In the future, the
subleading 3N contributions as well as consistently evolved 3N forces
up to N$^3$LO can be treated in the present framework once reliable
fits for the couplings $c_D$, $c_E$ are available. So far, fits based
on present NN interactions lead to unnaturally large $c_D$, $c_E$
couplings at N$^3$LO~\cite{Gola14n3lo}. Work in this direction is
currently in progress.

Our calculations are based on a perturbative expansion of the energy
up to second order around the Hartree-Fock state. In the Hartree-Fock
approximation, the energy density of isospin-asymmetric matter is
given by
\begin{align}
&\frac{E^{(1)}_{\rm NN} + E^{(1)}_{\overline{\rm 3N}}}{V} = \frac{1}{4\pi^3} \int dp \, p^2 \int dP \, P^2  \nonumber \\
&\times \sum_{m_T} \, f_{m_T}(\vec{p},\vec{P}) \sum_{L,S,J,T} (2J+1) \left(1 - (-1)^{L+S+T} \right) \nonumber \\ 
&\times \braket{p(LS)JTm_T|V_{\rm{NN}} + \overline{V}_{\rm{3N}}^{\rm{as}}
(P)/9|{p(LS)JTm_T}} \,,
\end{align}
with the short-hand notation $i=\vec{k}_i\sigma_i\tau_i$ and the
combinatorial factor~($1/9$) of the effective interaction
$\overline{V}_{\rm{3N}}^{\rm{as}}$ is discussed in detail in
Ref.~\cite{Hebe10nmatt}. Note that since the matrix elements in
Eq.~\eqref{eq:TBF_ME} involve two instead of one antisymmetrizer, a
relative conversion factor of $3$ is required for the comparison to
Ref.~\cite{Hebe10nmatt}. Furthermore, we have introduced the function
$f_{m_T}(\vec{p},\vec{P}) = \int d\cos \theta_{\vec{P},\vec{p}} \,
n_{\vec{P}/2+\vec{p}}^{\tau_1} \, n_{\vec{P}/2-\vec{p}}^{\tau_2}$,
which depends only on the two-body isospin projection quantum number
$m_T=\tau_1+\tau_2$ because the integrand is symmetric in the isospin
indices $\tau_1$ and $\tau_2$. It is important to constrain the
phase-space integral to the non-vanishing region of the Fermi-Dirac
distributions. The general case of the phase-space integral can be
written as
\begin{align}
I&= \int_{-1}^{+1} d\cos\theta_{\vec{p},\vec{P}} \; n_{\vec{p}+ \vec{P}/2}^{\tau_1} \, n_{\vec{p}- \vec{P}/2}^{\tau_2} \; f(\cos(\theta_{\vec{p},\vec{P}})) \nonumber \\
&= \Theta(x_\textrm{max}-x_\textrm{min}) \int_{x_\textrm{min}}^{x_\textrm{max}} d\cos \theta_{\vec{p},\vec{P}} \; f(\cos(\theta_{\vec{p},\vec{P}})) \, . 
\end{align}
In terms of $D_i^\pm(p,P) \equiv (k_{F,\tau_i}^2-P^2/4-p^2)/(\pm pP)$,
we obtain the limits,
\begin{subequations}
\begin{align}
x_\textrm{min}&=\max \left[-1.0, \min[+1.0,D_2^-(p,P) ]\right] \, , \\[1mm]
x_\textrm{max}&= \min\left[+1.0, \max[-1.0,D_1^+(p,P)]\right] \, .
\end{align}
\end{subequations}
Since $f(\cos(\theta_{\vec{p},\vec{P}})) = 1$ at the Hartree-Fock
level, this leads to $f_{m_T}(p,P) = \left(x_\textrm{max} -
x_\textrm{min}\right) \Theta(x_\textrm{max}-x_\textrm{min})$.

The second-order contribution to the energy density is given by
\begin{align}\label{eq:E_2nd}
\frac{E^{(2)}_{\rm NN} + E^{(2)}_{\overline{\rm 3N}}}{V} &= \frac{1}{4^2} \prod_{i=1}^{4} \left[ \text{Tr}_{\sigma_i} \text{Tr}_{\tau_i} \int \frac{d\vec{k}_i}{(2\pi)^3} \right] \bigl|\braket{12| V_{\rm as}^{(2)} | 34} \bigr|^2 \nonumber \\[1mm]
&\quad \times \frac{n_{\vec{k}_1}^{\tau_1} n_{\vec{k}_2}^{\tau_2} (1-n_{\vec{k}_3}^{\tau_3}) (1-n_{\vec{k}_4}^{\tau_4})}{\varepsilon_{\mathbf{k}_1}^{\tau_1}+\varepsilon_{\mathbf{k}_2}^{\tau_2}-\varepsilon_{\mathbf{k}_3}^{\tau_3}-\varepsilon_{\mathbf{k}_4}^{\tau_4}} \nonumber \\[1mm]
&\quad \times (2 \pi)^3 \delta(\vec{k}_1 + \vec{k}_2 - \vec{k}_3 -\vec{k}_4 ) \,.
\end{align}
Expanding in partial waves and performing the spin sums leads to (see Refs.~\cite{Tolo08nmatt,Hebe10nmatt})
\begin{align}
& \sum \limits_{S,S' M_S, M_{S'}}  \braket{\vec{p}S M_S T M_T| V_{\rm as}^{(2)}
|\vec{p}'S'  M_{S'} T' M_T} \nonumber \\
& \times \braket{\vec{p'}S'  M_{S'} T'' M_T|V_{\rm as}^{(2)} |\vec{p}S M_S T''' M_T} \nonumber \\
&=(4\pi)^2 \sum \limits_{S,S'} (-1)^{S+S'} \sum \limits_{\bar{L}} P_{\bar{L}}(\cos \theta_{\vec{p},\vec{p}'}) \nonumber \\
&\times \sum \limits_{L,L', \tilde{L},\tilde{L}'} \sum \limits_{J,\tilde{J}} i^{L-L'-\tilde{L}+\tilde{L}'} (-1)^{\bar{L}+\tilde{L}'+L'} \nonumber \\
& \times  \clebschG{L'}{0}{\tilde{L}'}{0}{\bar{L}}{0}\clebschG{L}{0}{\tilde{L}}{0}{\bar{L}}{0} \sixJsymbsixA{L}{\tilde{J}}{S}{\bar{L}}{J}{\tilde{L}} \sixJsymbsixA{\tilde{L}'}{J}{S'}{\bar{L}}{\tilde{J}}{\bar{L}'} \nonumber \\
&\times(2J+1)(2\tilde{J}+1) \, \sqrt{(2L+1)(2L'+1)(2\tilde{L}+1)(2\tilde{L}'+1)} \nonumber \\[1mm]
& \times  \braket{k'(L'S')JT'' M_{T}|V_{\rm as}^{(2)}|k(LS)JT''' M_T} \nonumber \\[1mm]
&\times \braket{k(\tilde{L}S)\tilde{J}T M_T| V_{\rm as}^{(2)}|k'(\tilde{L}'S')\tilde{J}T' M_{T}} \nonumber \\[1mm]
& \times  \left[1-(-1)^{\tilde{L}+S+T}\right] \left[1-(-1)^{L'+S'+T''}\right] \nonumber \\
& \times  \left[1-(-1)^{\tilde{L}'+S'+T'}\right] \left[1-(-1)^{L+S+T'''}\right] \, .
\end{align}
Here, the partial-wave interaction matrix elements are given by
$V_{\rm as}^{(2)} = V_\text{NN} + \overline{V}^{\text{as}}_{\rm{3N}} (P)/3$ (see
Ref.~\cite{Hebe10nmatt}), resulting from the normal-ordered two-body
part of 3N forces. $\{ \ldots \}$ denote $6j$-symbols and $P_L(\cos
\theta)$ are Legendre polynomials. The sums over the single-particle
isospin quantum numbers have to be performed explicitly, because the
Fermi-Dirac distribution functions break the isospin symmetry for
asymmetric matter. We stress that in general the effective interaction
$\overline{V}_{\rm{3N}}$ couples different spin and isospin channels
because of the isospin dependence of the Fermi-Dirac distribution
functions in Eq.~\eqref{eq:normord_jacobi}.

For the evaluation of Eq.~\eqref{eq:E_2nd} we need to calculate the
single-particle energies $\varepsilon_{\mathbf{k}}^{\tau}$, which are
in general determined by the solution of the Dyson equation
$\varepsilon_{\mathbf{k}}^{\tau} = k^2/(2m) + \text{Re}\,
\Sigma^{\tau} (k,\varepsilon_{\mathbf{k}})$. For our calculations, we
either use a free spectrum or compute the self-energy in the
Hartree-Fock approximation and average over the external spin quantum
numbers. Moreover, we average the isospin dependence, weighted by the
proton fraction $x$,
\begin{align}
& \Sigma^{(1)} (k_1,x) = \frac{1}{2 \pi} \int dk_2 \, k_2^2 \int d\cos \theta_{\vec{k}_1 \vec{k}_2}  \nonumber \\
&\times  \sum \limits_{T,M_T,\tau_1,\tau_2} w_{\tau_1} (x) \, n_{\vec{k}_2}^{\tau_2} \left( \clebschG{1/2}{\tau_1}{1/2}{\tau_2}{T}{M_T} \right)^2   \nonumber \\ 
&\times \sum \limits_{J,S,L} (2J+1) \, \left(1 - (-1)^{L+S+T} \right) \nonumber \\
&\times \braket{k_{12}/2 (L S) J T M_T | V_\text{NN} + \overline{V}^{\text{as}}_{\rm 3N}(P)/6 | k_{12}/2 (L S) J T M_T} \, ,
\end{align}
with $k_{12} = |\mathbf{k}_1 - \mathbf{k}_2|$ and the combinatorial
factor~($1/6$) of the effective interaction
$\overline{V}_{\rm{3N}}^{\rm{as}}$ is discussed in
Ref.~\cite{Hebe10nmatt}. The isospin weighting factor $w_{\tau}$ is
given by
\begin{equation}
w_\tau (x) =
\begin{cases}
x & \tau = +\frac{1}{2} \; \text{(proton)} \, ,\\
1-x & \tau = -\frac{1}{2} \; \text{(neutron)} \, .
\end{cases}
\end{equation}

In this approximation the single-particle energies for a certain
proton fraction $x$ are then $\varepsilon (k,x) = k^2/(2m) +
\Sigma^{(1)}(k,x)$, with $m$ being the average nucleon mass. In case
of the free spectrum, we apply only the kinetic energy as
single-particle energy. In neutron and symmetric matter, the isospin
weighting is equivalent to the ones in
Ref.~\cite{Hebe10nmatt,Hebe11fits} but includes also charge-symmetry
breaking.

\section{\label{sec:results}Results}

\subsection{\label{sec:PW_conv}Partial-wave convergence}

For our practical calculations we include 3N matrix elements up to
$\mathcal{J} = 9/2$ for the calculation of the effective interaction
$\overline{V}_{\rm{3N}}$ via Eq.~\eqref{eq:V_eff_pw}, where the 3N
matrix elements are calculated in the framework of
Ref.~\cite{Hebe15N3LOpw}. We have checked that this basis space leads
to well converged results for the effective NN potential up to
partial-wave channels with $J \lesssim 4$. In addition, we find
excellent agreement with the matrix elements of
$\overline{V}_{\text{3N}}$ at $P=0$ of Ref.~\cite{Hebe10nmatt} for
neutron matter and with the corresponding results for symmetric
nuclear matter~\cite{Hebe11fits} based on chiral 3N interactions at
N$^2$LO.

\begin{figure}[t]
\includegraphics[scale=1.0,clip=]{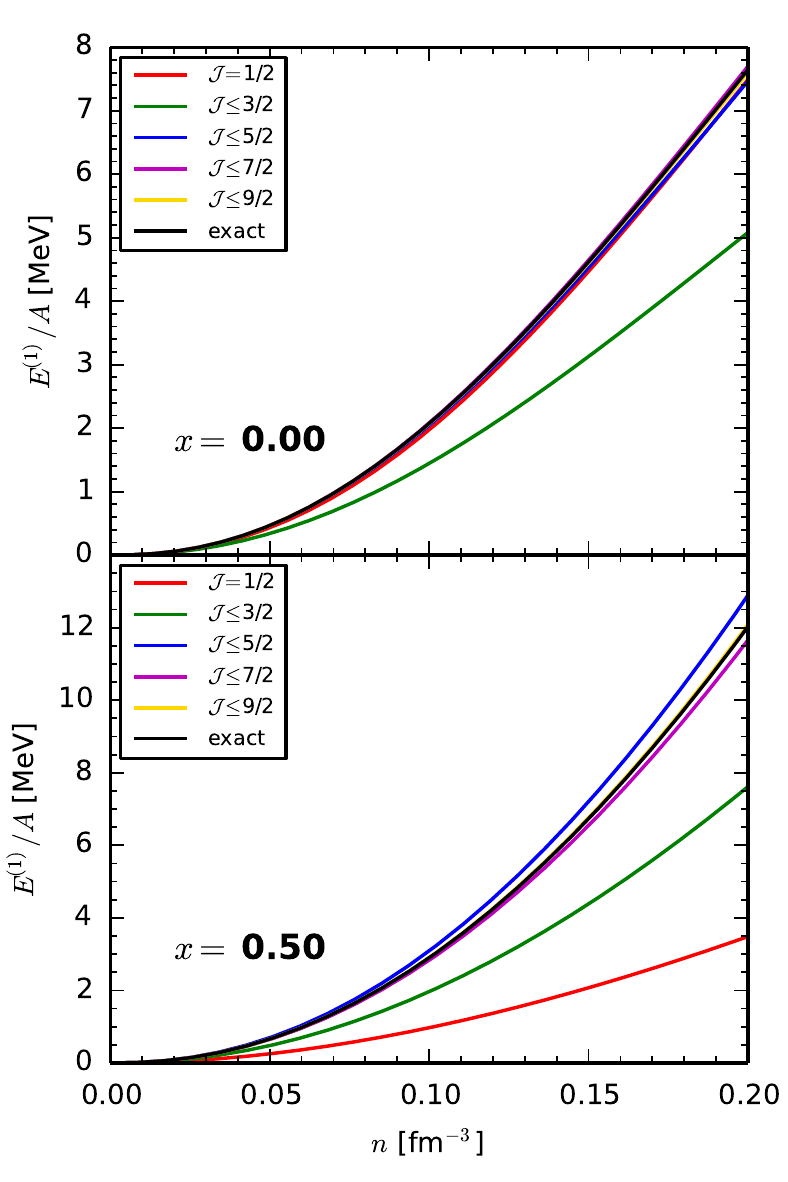}
\caption{\label{fig:3N_conv}(Color online)
Partial-wave convergence of the N$^2$LO 3N contributions at the
Hartree-Fock level in neutron matter (top) and symmetric matter
(bottom) for Hamiltonian~2 of Table~\ref{tab:couplings}.  For neutron
matter the contributions for \mbox{$\mathcal{J} > 5/2$} are very
small, so the individual lines are nearly indistinguishable.}
\end{figure}

As an additional benchmark we compare in Fig.~\ref{fig:3N_conv} the
Hartree-Fock contributions of 3N forces to the energy per particle
based on a summation of 3N matrix elements using different truncations
in $\mathcal{J}$ (following Ref.~\cite{Hebe13NMevol}) with results
derived directly from evaluating the operatorial structure of the
N$^2$LO 3N interactions (see Refs.~\cite{Tolo08nmatt,Hebe10nmatt}) for
neutron matter (top panel) and for symmetric nuclear matter (lower
panel). These two independent calculations test directly the
convergence of the partial-wave decomposition and should provide
identical results in the limit $\mathcal{J} \leqslant
\mathcal{J}_{\rm{max}} \rightarrow \infty$ up to numerical uncertainties.

The results shown in Fig.~\ref{fig:3N_conv} are based on the set of
low-energy couplings of Hamiltonian~2 of Table~\ref{tab:couplings} and
by including all contributions with $J \leqslant 6$ for each 3N
partial wave. We find excellent agreement of the results for
\mbox{$\mathcal{J} \leqslant 9/2$}, with a deviation of less than
100~keV at saturation density for neutron matter and symmetric nuclear
matter.  Hence, for the following we will use this basis space for the
calculation of the effective interaction
$\overline{V}_{\rm{3N}}^{\rm{as}}$.

\subsection{\label{sec:Disc_EOS}Discussion of the equation of state}

\begin{figure*}
\includegraphics[page=1,scale=1.0,clip=]{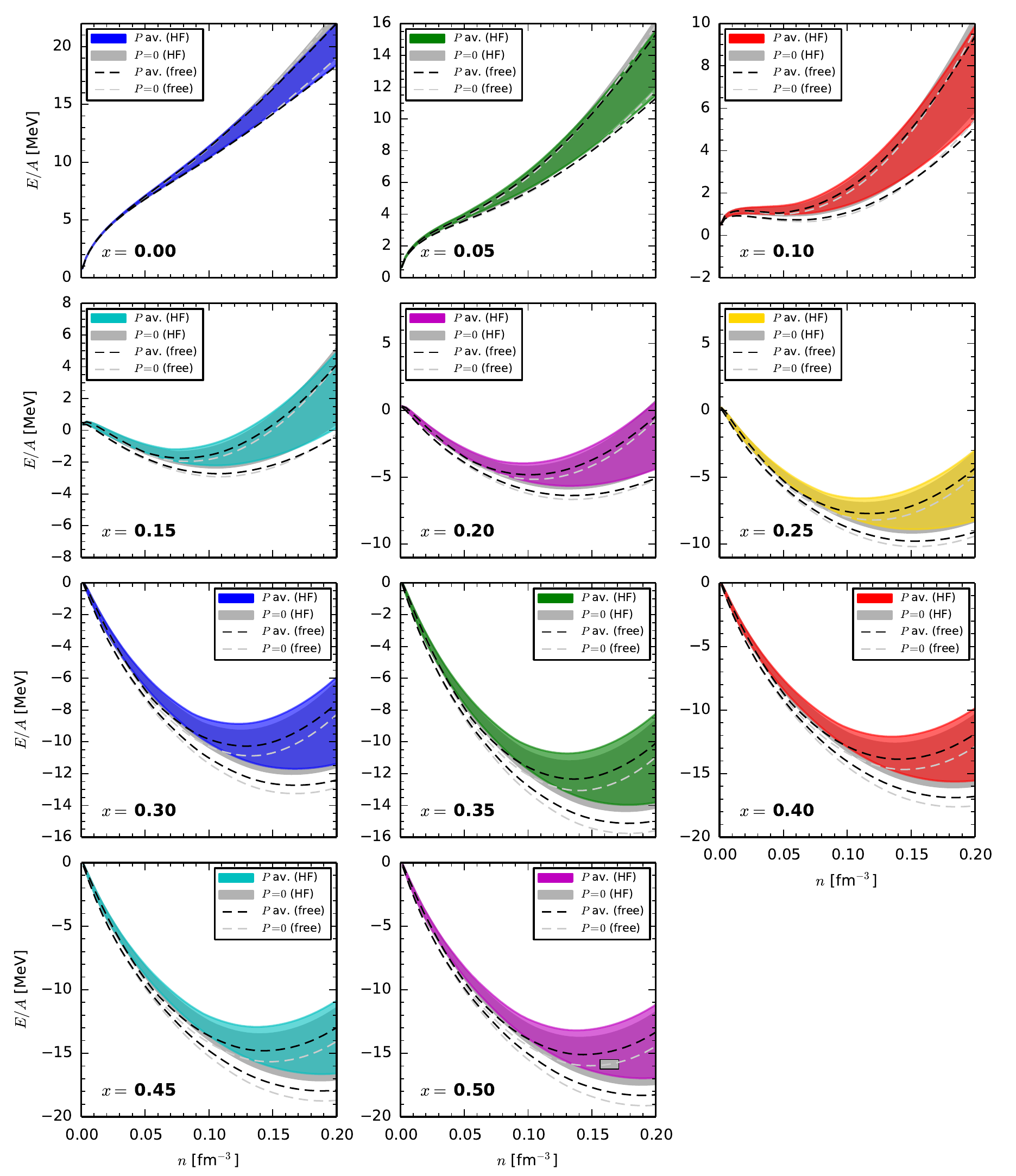}
\caption{\label{fig:EOS_panel}(Color online)
Energy per particle of nuclear matter as a function of the total
density $n = n_n + n_p$ for various proton fractions $x$. The two
approximations of the effective NN potential, $P = 0$ and $P$ average,
and two approximations for the single-particle energies, free and
Hartree-Fock, are shown. The energy range is based on the set of
Hamiltonians listed in Table~\ref{tab:couplings}. The excluded
Hamiltonian~6* has no influence on the uncertainty bands. For a better
view, the area between the dashed lines are not filled in the case of
a free spectrum. For $x=0.5$ we also show the empirical saturation
point (see text for details).}
\end{figure*}

In Fig.~\ref{fig:EOS_panel}, we show the results for the energy per
particle for eleven proton fractions using different approximations
for the single-particle energies and the effective interactions
$\overline{V}_{\rm{3N}}$: the dashed lines show the results based on
the free single-particle spectrum (i.e., \mbox{$\Sigma^{(1)} (k_1,x) =
0$}), whereas the solid lines show the results based on the
single-particle energies calculated in the Hartree-Fock approximation;
the colored/gray bands represent results based on the effective NN
potential calculated in the $P$-averaged/$P=0$ approximation,
respectively. For each of these four sets of results we determine the
theoretical uncertainties by performing calculations based on the
Hamiltonians listed in Table~\ref{tab:couplings} and extract the
maximal spread of these results.  We note that for Hamiltonian~6* the
fits of the short-range 3N couplings $c_D, c_E$ in
Ref.~\cite{Hebe11fits} have not taken into account the
isospin-breaking of the N$^3$LO NN potential EGM~550/600~MeV. This
leads to deviations for the $^3$H binding energy of $\sim 200$~keV in
comparison to exact calculations. We will discuss this Hamiltonian
separately, but emphasize that it has no influence on the uncertainty
bands shown in Fig.~\ref{fig:EOS_panel}.

From the many-body point of view, neutron matter~($x=0$) represents
the simplest system. At N$^2$LO, only the long-range 3N forces
proportional to $c_1$ and $c_3$ contribute for non-local regulators
$f_\text{R}(p,q)$ due to the Pauli-principle and the isospin structure
of the 3N forces~\cite{Hebe10nmatt}.  In addition, no NN $S$-wave
tensor interactions are active in neutron matter.  As a results we
find relatively narrow uncertainty bands with a width of about $4$~MeV
at saturation density.

Increasing the proton fraction influences the overall uncertainty in
two ways.  First, the width of the bands for each of the two
single-particle spectra becomes larger up to $~(5-6)$~MeV for
symmetric nuclear matter at the highest density shown. The upper
uncertainty limit is always determined by Hamiltonian~7, which also
leads to rather small saturation densities for $x=0.5$~(see
below). Second, the difference between the individual results based on
the two spectra grows systematically for larger proton fractions.  The
dependence of our results on the single-particle energies probes the
perturbativeness of the Hamiltonians and provides a measure of
contributions from higher orders in the perturbative expansion. Hence,
these results indicate the need to analyze third-order contributions
more closely, which we discuss in Sec.~\ref{sec:trd}. The two
approximations lead to comparable results and widely overlapping bands
in each spectrum. We find that the $P$-average approximation is
slightly more repulsive, and the bands are shifted at saturation
density by $\approx 1$~MeV in symmetric matter.

\begin{figure}[t]
\includegraphics[scale=1.0,clip=]{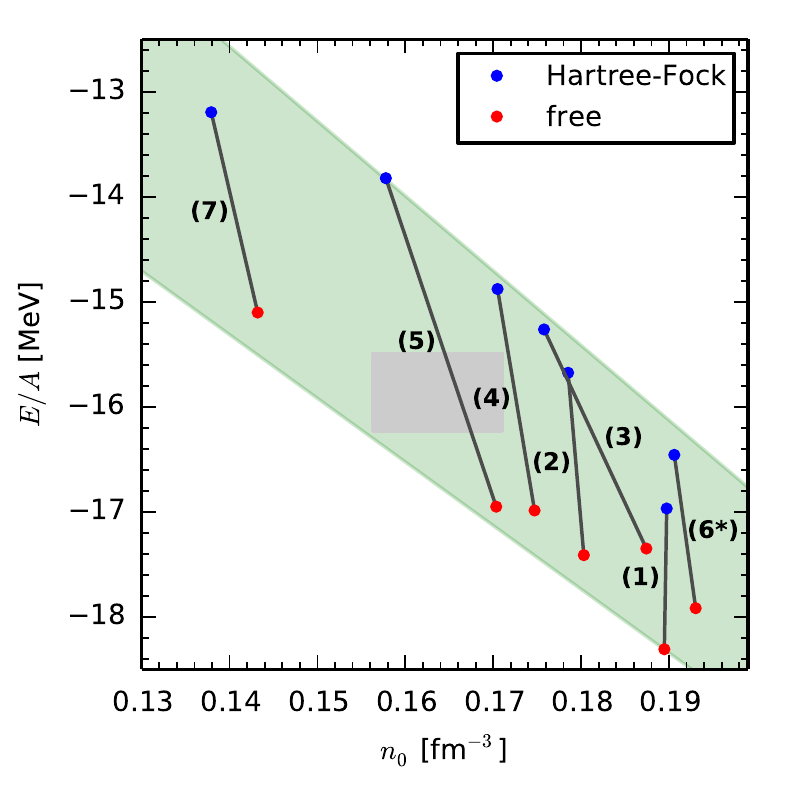}
\caption{\label{fig:Coester}(Color online) 
Correlation between the saturation density and energy for the seven 
Hamiltonians of Table~\ref{tab:couplings}, indicated by the
figure. The green area highlights the obtained Coester band based on
independently fitting the saturation points for a free and a
Hartree-Fock spectrum. The gray lines connect the two spectra.  As
discussed in the text, the empirical saturation point (gray box) is
given by the range of 14 selected energy-density functionals. The
region is in good agreement with our calculated Coester band. See the
text for details of Hamiltonian~6*.}
\end{figure}

We discuss now the properties of the equation of state based on the
$P$-average approximation of the effective interaction. Considering
the free and the Hartree-Fock spectrum and excluding Hamiltonian~6*,
symmetric matter saturates at \mbox{$n_0=(0.138-0.190)$~fm$^{-3}$}
with energies of \mbox{$E/A=-(13.2-18.3)$~MeV}. Hamiltonian~6*
increases slightly the upper limit by \mbox{$\Delta
n_0=0.003$~fm$^{-3}$}. The saturation points for each of the seven
Hamiltonians of Table~\ref{tab:couplings} are shown in
Fig.~\ref{fig:Coester}. The red (blue) points correspond to the
calculations with a free (Hartree-Fock) spectrum. Hence, the gray line
connecting the two calculations indicates the convergence of the
calculation.  We find a Coester-like linear correlation between the
energy and density at the saturation point, but the range is
considerably smaller than the Coester line based on NN interactions
only~\cite{Coes70Var}.  The green band has been obtained by
independently fitting a linear function to the saturation points for
the two spectra excluding Hamiltonian~6*.

Skyrme energy-density functionals based on properties of nuclei and
nuclear matter can be used to empirically constrain the saturation
point~\cite{Dutr12SkNM,Brow13skyrme,Brow14skyrme}. Table~7 of
Ref.~\cite{Dutr12SkNM} summarizes 16 selected functionals, which
reproduce well selected properties of nuclear matter. Six more are
excluded because of unreasonable behavior for large
densities~\cite{Dutr12SkNM} or being unstable for finite nuclei. The
remaining ten are listed in Table~1 of Ref.~\cite{Brow14skyrme}. Our
empirical saturation range is determined based on these functionals
plus those of Ref.~\cite{Kort14opt} (SLy4, UNEDF0, UNEDF1, and,
UNEDF2). As a result we obtain the ranges \mbox{$n_0^\text{emp} =
(0.164\pm 0.007)~\text{fm}^{-3}$} and \mbox{$E^\text{emp}/A  \simeq
-(15.9 \pm 0.4)~\text{MeV}$}, which is indicated by the
gray boxes in Fig.~\ref{fig:EOS_panel} (for $x=0.5$) and
Fig.~\ref{fig:Coester}.

Our band in Fig.~\ref{fig:Coester}, based on NN and 3N interactions,
overlaps with the empirical saturation point, in contrast to
calculations based on NN interactions only~\cite{Coes70Var}. This
holds especially for the equation of state based on Hamiltonian~4 and
5. We note that Hamiltonian~5 has the largest dependence on the
spectrum as it is almost twice compared to Hamiltonian~1.  This may be
due to the large resolution scale $\lambda=2.8$~fm$^{-1}$ of the
Hamiltonian. Although the $^3$H binding energy corresponding to
Hamiltonian~6* is not well fit, it behaves still natural and similar
as Hamiltonian~1.

\begin{table*}[t]
\caption{\label{tab:coeff}
Coefficients $C_{\mu\nu}$ of the quadratic expansion~\eqref{eq:fit} fit 
to the calculated equation of state $E/A(\beta,\bar{n})$ for each
Hamiltonian. The values are from separately fitting neutron and
symmetric nuclear and then extending quadratically in $\beta$
according to Eq.~\eqref{eq:fit}. The coefficients are given in MeV.}
\begin{ruledtabular}
\begin{tabular}{lrrrrrrrrrr}
& \multicolumn{1}{c}{$C_{02}$} & \multicolumn{1}{c}{$C_{03}$} & \multicolumn{1}{c}{$C_{04}$} & \multicolumn{1}{c}{$C_{05}$} & \multicolumn{1}{c}{$C_{06}$} & \multicolumn{1}{c}{$C_{22}$} & \multicolumn{1}{c}{$C_{23}$} & \multicolumn{1}{c}{$C_{24}$} & \multicolumn{1}{c}{$C_{25}$} & \multicolumn{1}{c}{$C_{26}$}  \\ 
 \hline
\#~1  & $0.3$ & $-66.1$ & $79.2$ & $-50.3$ & $20.3$ & $6.1$ & $156.2$ & $-306.0$ & $259.3$ & $-83.8$ \\ 
\#~2  & $4.1$ & $-78.2$ & $92.5$ & $-53.7$ & $19.7$ & $2.3$ & $168.5$ & $-319.9$ & $263.8$ & $-83.6$ \\ 
\#~3  & $12.5$ & $-125.2$ & $191.0$ & $-141.4$ & $48.0$ & $4.1$ & $153.0$ & $-275.9$ & $209.2$ & $-59.7$ \\ 
\#~4  & $6.5$ & $-83.0$ & $92.8$ & $-47.7$  & $16.6$ & $-0.2$ & $174.5$ & $-322.6$ & $260.2$ & $-81.2$ \\ 
\#~5  & $9.1$ & $-78.9$ & $65.0$ & -$13.2$  & $4.2$ & $-4.0$ & $178.3$ & $-311.7$ & $240.0$ & $-72.6$ \\ 
\#~6* & $2.2$ & $-77.0$ & $106.6$ & $-76.7$ & $28.8$ & $-4.8$ & $219.0$ & $-440.3$ & $377.1$ & $-119.6$ \\ 
\#~7  & $-0.9$ & $-54.1$ & $52.9$ & $-28.5$ & $17.6$ & $-1.6$ & $194.8$ & $-385.0$ & $332.9$ & $-110.6$ \\ 
\end{tabular}
\end{ruledtabular}
\end{table*}

Following the usual quadratic expansion in the isospin asymmetry
$\beta$, we approximate globally the equation of state in terms of a
power series in the reduced density $\bar{n}=n/(0.16\,\text{fm}^{-3})$
\begin{equation}\label{eq:fit}
\frac{E}{A}(\beta,\bar{n}) = \sum \limits_{\substack{\mu=0,2\\ \nu=2,3,4,5,6}} C_{\mu\nu} \, \beta^\mu \, \bar{n}^{\nu/3} \, .
\end{equation}
In order to determine the coefficients $C_{\mu\nu}$, we fit to the
energy per particle of neutron and symmetric nuclear matter and
interpolate then quadratically in $\beta$ to general isospin
asymmetries. We constrain the fit to densities of
\mbox{$(0.02-0.22)~\text{fm}^{-3}$}. The resulting values of the
coefficients are listed in Table~\ref{tab:coeff}.  Based on the
root-mean-square deviation of the global energy expression and the
equation of state of each calculated asymmetry, we have checked that
Eq.~\eqref{eq:fit} provides a reasonable approximation for our
microscopic results, especially close to symmetric matter. At
$n=0.16$~fm$^{-3}$, the largest deviation to the data is $\lesssim
220$~keV for neutron-rich matter, while the typical value for larger
proton fractions is much smaller. We stress that these coefficients
only represent results from a least-squares fit to our data, and are
given here just for completeness. In particular, any physical
interpretation of these coefficients has to be done with care due to
the large values of some coefficients (see Table~\ref{tab:coeff}) and
the resulting cancellations of terms.

Using the parametrization~\eqref{eq:fit} we compute the
incompressibility $K$ and the symmetry energy $S_v$,
\begin{subequations} \label{eq:ana_deriv}
\begin{align} 
K = 9\, \frac{\partial^2}{\partial \bar{n}^2} \frac{E}{A}(\beta,\bar{n}) \bigg|_{\substack{\bar{n}=1\\ \beta=0}} \; ,\\
S_v = \frac{1}{2} \frac{\partial^2}{\partial \beta^2} \frac{E}{A}(\beta,\bar{n}) \bigg|_{\substack{\bar{n}=1\\ \beta=0}} \; ,
\end{align}
\end{subequations}
at the actual saturation density of each Hamiltonian. Based on the
uncertainty ranges of our results (shown as the colored bands in
Fig.~\ref{fig:EOS_panel}) we obtain the ranges
\mbox{$K=(182-254)$~MeV} and \mbox{$S_v=(28.4-35.7)$~MeV}, considering
the free and the Hartree-Fock spectrum and excluding
Hamiltonian~6*. Hamiltonian~6* would increase the upper uncertainty
limits to \mbox{$\Delta K = 8$~MeV} and \mbox{$\Delta S_v =
0.4$~MeV}. We note that the value of \mbox{$S_v=(30.2-32.2)$~MeV}
(at fixed density \mbox{$n_0=0.16$~fm$^{-3}$}), which we obtained in
Ref.~\cite{Dris14asymmat} for small proton fractions, is in agreement
with these improved calculations. The uncertainty here is larger
because Eqs.~\eqref{eq:ana_deriv} are evaluated here at the actual
saturation density of the Hamiltonian and not at fixed
\mbox{$n_0=0.16$~fm$^{-3}$}.  We will study the properties of the
symmetry energy and the importance of a quartic term~($\sim \beta^4$)
of the energy expansion in a subsequent paper.

\subsection{\label{sec:trd}Estimate of the third-order contribution}

\begin{figure}[t]
\includegraphics[scale=1.0,clip=]{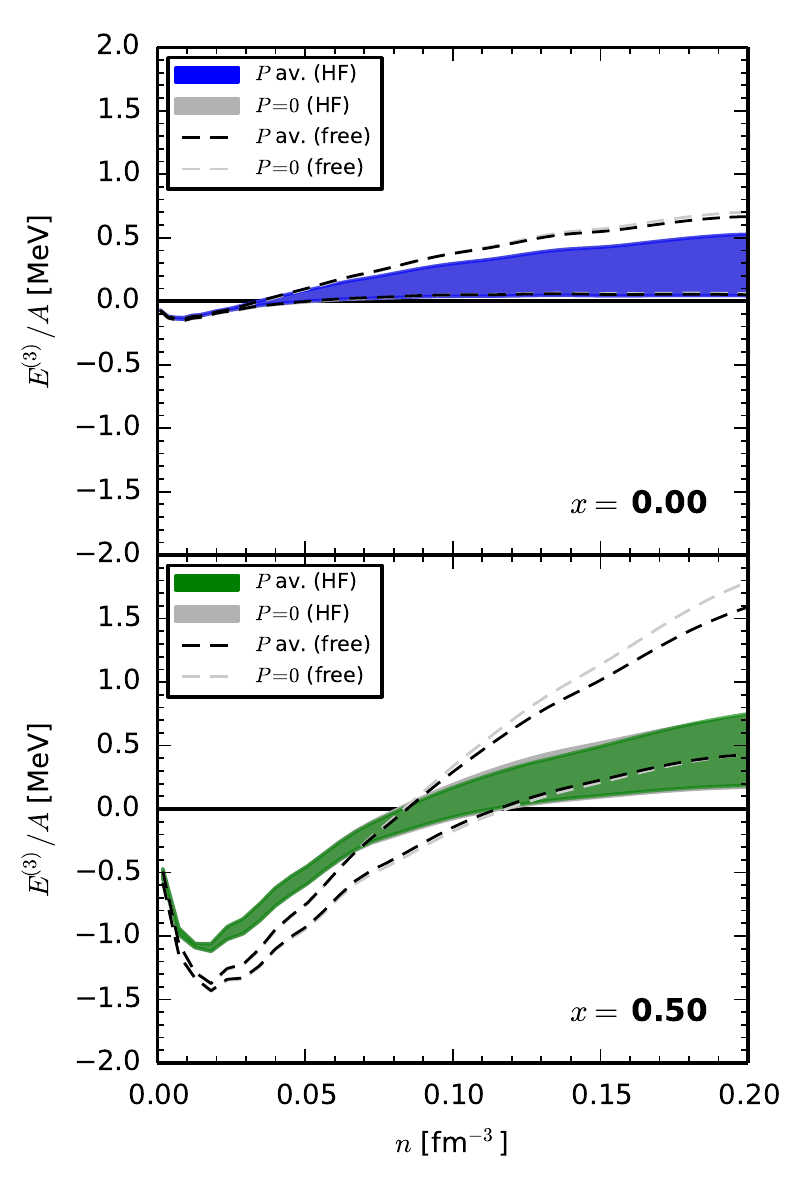} 
\caption{\label{fig:trd}(Color online)
Contributions at third order in the perturbative expansion to the
energy of neutron matter (top) and symmetric nuclear matter
(lower panel). The color coding and line styles are as in
Fig.~\ref{fig:EOS_panel}. Accordingly, the uncertainty estimates are
based the Hamiltonians of Table~\ref{tab:couplings}.}
\end{figure}

The results shown in Fig.~\ref{fig:EOS_panel} exhibit a mild
sensitivity to the single-particle spectrum employed, which indicates
that contributions beyond second order in the perturbative expansion
might give non-negligible contributions.  Here, we estimate
third-order contributions to neutron and symmetric nuclear matter in
order to assess the quality of the perturbative convergence. At third
order, the diagrams involve particle-particle, hole-hole and
particle-hole excitations. We consider here only particle-particle and
hole-hole diagrams, respectively (see Ref.~\cite{Hebe11fits} for
details).  The calculations are simplified by employing angle-averaged
Fermi-Dirac distribution functions. Figure~\ref{fig:trd} shows the
corresponding third-order contributions, $E^{(3)}/A$, in neutron~(top)
and symmetric matter~(bottom panel) for the two approximations of the
effective interaction and using free/Hartree-Fock single-particle
energies.

At $n_0=0.16$~fm$^{-3}$ and using a Hartree-Fock (free) spectrum, we
find repulsive contributions of up to $\sim 500$~keV ($\sim 600$~keV)
in neutron matter and $\sim(100-600)$~keV [$\sim(0.3-1.3)$~MeV] in
symmetric matter. The NN interaction dominates the overall
contributions. However, 3N forces become more important for larger
proton fractions. These findings are consistent with the results in
Ref.~\cite{Bogn05nuclmat}, based on low-momentum interactions
$V_{\text{low }k}$. In future work, we will study the order-by-order
convergence in the many-body expansion by including also particle-hole
contributions (see also Ref.~\cite{Cora14MBPT}). We emphasize that
Hamiltonian~6* (see Table~\ref{tab:couplings}) does not influence
these uncertainty estimates.

\begin{figure*}
\includegraphics[page=1,scale=1.0,clip=]{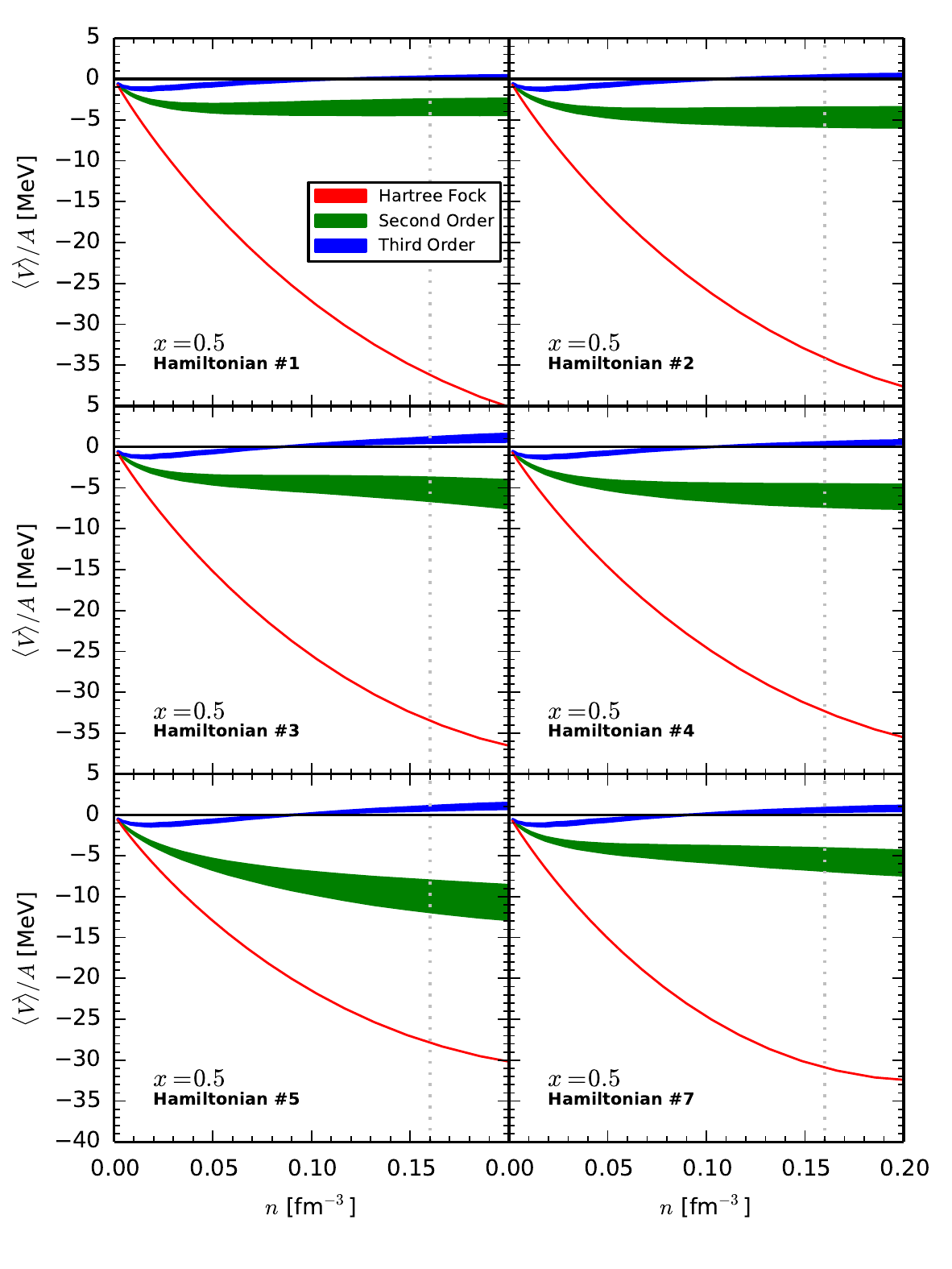}
\caption{\label{fig:EOS_contr}(Color online) Interaction energy at first, 
second and third order in symmetric nuclear matter ($x=0.5$) for the
employed Hamiltonians of Tab.~\ref{tab:couplings}. As discussed in the
text Hamiltonian 6* is excluded. The suppression of higher orders in
the perturbative expansion demonstrates the convergence of many-body
perturbation theory and suggests that the sensitivity of the
second-order results on the single-particle energies provide a
conservative estimate for the many-body uncertainties.}
\end{figure*}

In order to address the perturbative convergence we show in
Fig.~\ref{fig:EOS_contr} for each Hamiltonian of
Table~\ref{tab:couplings} the energy contributions $\braket{V}/A$ to
the free Fermi gas at first (red), second (green), and third order
(blue) in many-body perturbation theory. The variation of the two
single-particle spectra defines the uncertainty bands at second and
third order. The figure shows that the second-order results are
suppressed by a factor of $\sim 6$ compared to Hartree Fock whereas
the third-order estimates are suppressed by a factor of $\sim 5$
relative to second order at the largest density shown,
$n=0.2$~fm$^{-3}$. The band at second order is typically $\sim 10\%$
with respect to the first-order contribution. Only the Hamiltonian~(5)
with the large resolution scale $\lambda = 2.8$~fm$^{-1}$ shows, as
expected, a larger sensitivity on the single-particle spectrum and a
weaker suppression of higher-order terms in many-body perturbation
theory.

In conclusion, Fig.~\ref{fig:EOS_contr} demonstrates the convergence
of many-body perturbation theory for the employed Hamiltonians and
suggests that the sensitivity of the second-order results on the
single-particle energies provide a conservative estimate for the
many-body uncertainties.

\section{\label{sec:summary}Summary and outlook}

We have presented results for isospin-asymmetric matter based on
N$^3$LO NN and N$^2$LO 3N interactions calculated in many-body
perturbation theory. The contributions from three-body forces beyond
the Hartree-Fock approximation are included via an improved
normal-ordering framework. This novel framework is based on
partial-wave 3N matrix elements and makes it possible to generalize
the computation of the effective density-dependent two-body
interaction to finite center-of-mass momenta. In addition, it is also
straightforward to include contributions from subleading 3N
interactions at N$^3$LO~\cite{Bern083Nlong,Bern113Nshort} by utilizing
a new method for decomposing efficiently 3N interactions in a
partial-wave plane-wave basis~\cite{Hebe15N3LOpw}. Such full N$^3$LO
calculations can be performed immediately once reliable fits for the
low-energy couplings $c_D, c_E$ are available.

We employed the new normal-ordering framework to calculate the
effective potential as a function of the center-of-mass momentum
$\mathbf{P}$ by averaging this vector over all angles. We benchmarked
our results against previous results for vanishing $P$ and probed the
sensitivity of the energy per particle at the Hartree-Fock level to
different approximations in the normal ordering. We found that both
approximations, $P=0$ and $P$-averaging, provide good agreement with
exact results up to intermediate densities of about $n \sim 0.13 \,
\text{fm}^{-3}$, whereas the $P=0$ approximation becomes unreliable
beyond this density. In contrast, the new $P$-averaging approximation
remains stable and close to the calculated results for all relevant
densities.

For our many-body calculations we followed the strategy of
Ref.~\cite{Hebe11fits}. The NN forces were evolved via the SRG,
whereas the short-range couplings of the 3N interactions at N$^2$LO
were fit to few-body observables at a given NN resolution scale. The
theoretical uncertainties of our many-body observables are determined
by the range obtained from the different Hamiltonians listed in
Table~\ref{tab:couplings}. Recently, this has been successfully used
to study ab initio the charge radius, the neutron radius, the weak
form factor and the dipole polarizability of
$^{48}$Ca~\cite{Hage1548Ca}. In addition, we estimated the many-body
uncertainties by employing different approximations for the
normal-ordering of the 3N interactions and using different
approximations for the single-particle energies.

Based on the results for the energies at different proton fractions,
we fist calculated a global analytical fit in density and proton
fraction for each Hamiltonian and then extracted results for the
saturation point of symmetric matter, the incompressibility and for
the symmetry energy by using the standard quadratic expansion around
symmetric matter. We found a Coester-like linear correlation between
saturation density and energy, and the band covers the empirical
range. In addition, we found that a quadratic parameterization in the
isospin asymmetry reproduces the microscopic results reasonably well.

As next steps, it will be crucial to improve the estimates of the
theoretical uncertainties and also to investigate different regulator
choices.  As an example, in Refs.~\cite{Epel14improved,Epel14NNn4lo}
novel NN potentials at orders LO, NLO, N$^2$LO, N$^3$LO and N$^4$LO
and different regulator scales were derived. The present many-body
framework allows to perform systematic order-by-order convergence
studies in the chiral expansion at different regulator scales based on
such potentials, including 3N forces up to N$^3$LO.  In addition, the
present framework can be generalized by performing the normal-ordering
with respect to a general correlated reference state, by extending the
calculations to finite temperature, and by incorporating particle-hole
and higher-order contributions in the many-body expansion. This will
allow also systematic convergence studies in the many-body expansion.

\begin{acknowledgments}

We thank T. Kr{\"u}ger, J. M. Lattimer and I. Tews for useful
discussions. This work was supported by the ERC Grant No. 307986
STRONGINT.

\end{acknowledgments}

\hypersetup{
     colorlinks   = true,
     linkcolor = blue,
     citecolor = blue,
     menucolor = black,
     urlcolor = black  
}

\bibliography{literature}
\end{document}